\documentclass[prb, 11pt,letterpaper,superscriptaddress,floatfix,nofootinbib,notitlepage]{revtex4-1}

\usepackage{textcomp}
\usepackage{amsmath}
\usepackage{amsfonts}
\usepackage{amssymb}
\usepackage{graphicx}
\usepackage{siunitx}
\usepackage{caption}
\usepackage{setspace}
\setcitestyle{super}
\usepackage[colorlinks, citecolor={blue}, linkcolor={black}]{hyperref}

\begin{document}
\raggedbottom

\author{Yuan Cao}
\email{caoyuan@mit.edu}
\affiliation{Department of Physics, Massachusetts Institute of Technology, Cambridge, Massachusetts 02139, USA}
\author{Daniel Rodan-Legrain}
\affiliation{Department of Physics, Massachusetts Institute of Technology, Cambridge, Massachusetts 02139, USA}
\author{Oriol Rubies-Bigorda}
\affiliation{Department of Physics, Massachusetts Institute of Technology, Cambridge, Massachusetts 02139, USA}
\author{Jeong Min Park}
\affiliation{Department of Physics, Massachusetts Institute of Technology, Cambridge, Massachusetts 02139, USA}
\affiliation{Department of Chemistry, Massachusetts Institute of Technology, Cambridge, Massachusetts 02139, USA}
\author{Kenji Watanabe}
\author{Takashi Taniguchi}
\affiliation{National Institute for Materials Science, Namiki 1-1, Tsukuba, Ibaraki 305-0044, Japan}
\author{Pablo Jarillo-Herrero}
\email{pjarillo@mit.edu}
\affiliation{Department of Physics, Massachusetts Institute of Technology, Cambridge, Massachusetts 02139, USA}

\title{Electric Field Tunable Correlated States and Magnetic Phase Transitions in Twisted Bilayer-Bilayer Graphene}

\maketitle

\textbf{
The recent discovery of correlated insulator states and superconductivity in magic-angle twisted bilayer graphene \cite{cao_correlated_2018, cao_unconventional_2018} has paved the way to the experimental investigation of electronic correlations in tunable flat band systems realized in twisted van der Waals heterostructures \cite{suarez_morell_flat_2010, bistritzer_moire_2011,lopes_dos_santos_continuum_2012,carr_twistronics:_2017}. This novel twist angle degree of freedom and control should be generalizable to other 2D systems, which may exhibit similar correlated physics behavior while at the same time enabling new techniques to tune and control the strength of electron-electron interactions. Here, we report on a new highly tunable correlated system based on small-angle twisted bilayer-bilayer graphene (TBBG), consisting of two rotated sheets of Bernal-stacked bilayer graphene. We find that TBBG exhibits a rich phase diagram, with tunable correlated insulators states that are highly sensitive to both twist angle and to the application of an electric displacement field, the latter reflecting the inherent polarizability of Bernal-stacked bilayer graphene \cite{mccann_electronic_2013, castro_neto_electronic_2009}. We find correlated insulator states that can be switched on and off by the displacement field at all integer electron fillings of the moir\'{e} unit cell. The response of these correlated states to magnetic fields points towards evidence of electrically switchable magnetism. Moreover, the strong dependence of the resistance at low temperature near the correlated insulator states indicates possible proximity to a superconducting phase. Furthermore, in the regime of lower twist angles, TBBG shows multiple sets of flat bands near charge neutrality, resulting in numerous correlated states corresponding to half-filling of each of these flat bands. Our results pave the way to the exploration of novel twist-angle and electric-field controlled correlated phases of matter in novel multi-flat band twisted superlattices.
}

Electronic correlations play a fundamental role in condensed matter systems where the bandwidth is comparable to or less than the Coulomb energy between electrons. These correlation effects often manifest themselves as intriguing quantum phases of matter, such as a ferromagnetism, superconductivity, Mott insulators, or fractional quantum Hall states. Understanding, predicting and characterizing these correlated phases is of great interest in modern condensed matter physics research and poses challenges to both experimentalists and theorists. Recent studies in twisted graphene superlattices provide us with an ideal tunable platform to investigate electronic correlations in two dimensions\cite{cao_correlated_2018, cao_unconventional_2018, yankowitz_tuning_2019,sharpe_emergent_2019, lu_superconductors_2019}. Tuning the twist angle of 2D van der Waals heterostructures to realize novel electronic states, an emerging field referred to as `twistronics', has enabled physicists to explore a variety of novel phenomena. \cite{hunt_massive_2013,ponomarenko_cloning_2013, dean_hofstadters_2013,kumar_high-temperature_2017}. When two layers of graphene are twisted by a specific angle, the phase diagram in the system exhibits correlated insulator states with similarities to Mott insulator systems \cite{cao_correlated_2018}, as well as exotic superconducting states upon charge doping\cite{cao_unconventional_2018,yankowitz_tuning_2019, lu_superconductors_2019}. These effects likely stem from the many-body interactions between the electrons, when the band structure becomes substantially flattened as the twist angle approaches the first magic angle $\theta\approx\SI{1.1}{\degree}$\cite{suarez_morell_flat_2010, bistritzer_moire_2011,lopes_dos_santos_continuum_2012}.  

Here we extend the twistronics research on twisted bilayer graphene to a new system with electrical displacement field tunability --- twisted bilayer-bilayer graphene (TBBG), which consists of two sheets of untwisted Bernal-stacked bilayer graphene stacked together at an angle $\theta$, as illustrated in Fig. 1a. Compared to monolayer graphene, the band structure of bilayer graphene is highly sensitive to the applied perpendicular electric displacement field through it \cite{mccann_electronic_2013, oostinga_gate-induced_2008, zhang_direct_2009}, and therefore provides us with an extra knob to control the relative strength of electronic correlations in the bands\cite{chen_evidence_2019}. Similar to that of twisted bilayer graphene (TBG)\cite{suarez_morell_flat_2010,bistritzer_moire_2011,lopes_dos_santos_continuum_2012}, the band structure of TBBG is flattened near $\sim \SI{1.1}{\degree}$ (see Fig. 2e-g)\cite{zhang_nearly_2019}. For devices with twist angles near this value, our experiments show that the correlated insulator behavior at half-filling $n_s/2$ in TBBG can be sensitively turned on and off by the displacement field, where $n_s$ is the density corresponding to fully filling one spin and valley degenerate superlattice band\cite{cao_superlattice-induced_2016, kim_charge_2016}. The correlated insulator states are sensitive to the twist angle, and devices with slightly larger twist angles (e.g. \SI{1.23}{\degree}) exhibit a richer phase diagram, with multiple correlated insulating states at quarter fillings of the superlattice, $\frac{k}{4}n_s$, where $k$ is an odd number\cite{cao_unconventional_2018,yankowitz_tuning_2019,chen_signatures_2019, lu_superconductors_2019}. On the other hand, devices with smaller twist angles, such as \SI{0.84}{\degree}, display multiple correlated states at higher fillings, indicating the presence of multiple flat bands in the electronic structure. The combination of twist angle, electric displacement field and magnetic field control therefore provides a rich arena to investigate novel phenomena in the emerging field of twistronics. 

\begin{figure}
\includegraphics[width=\textwidth]{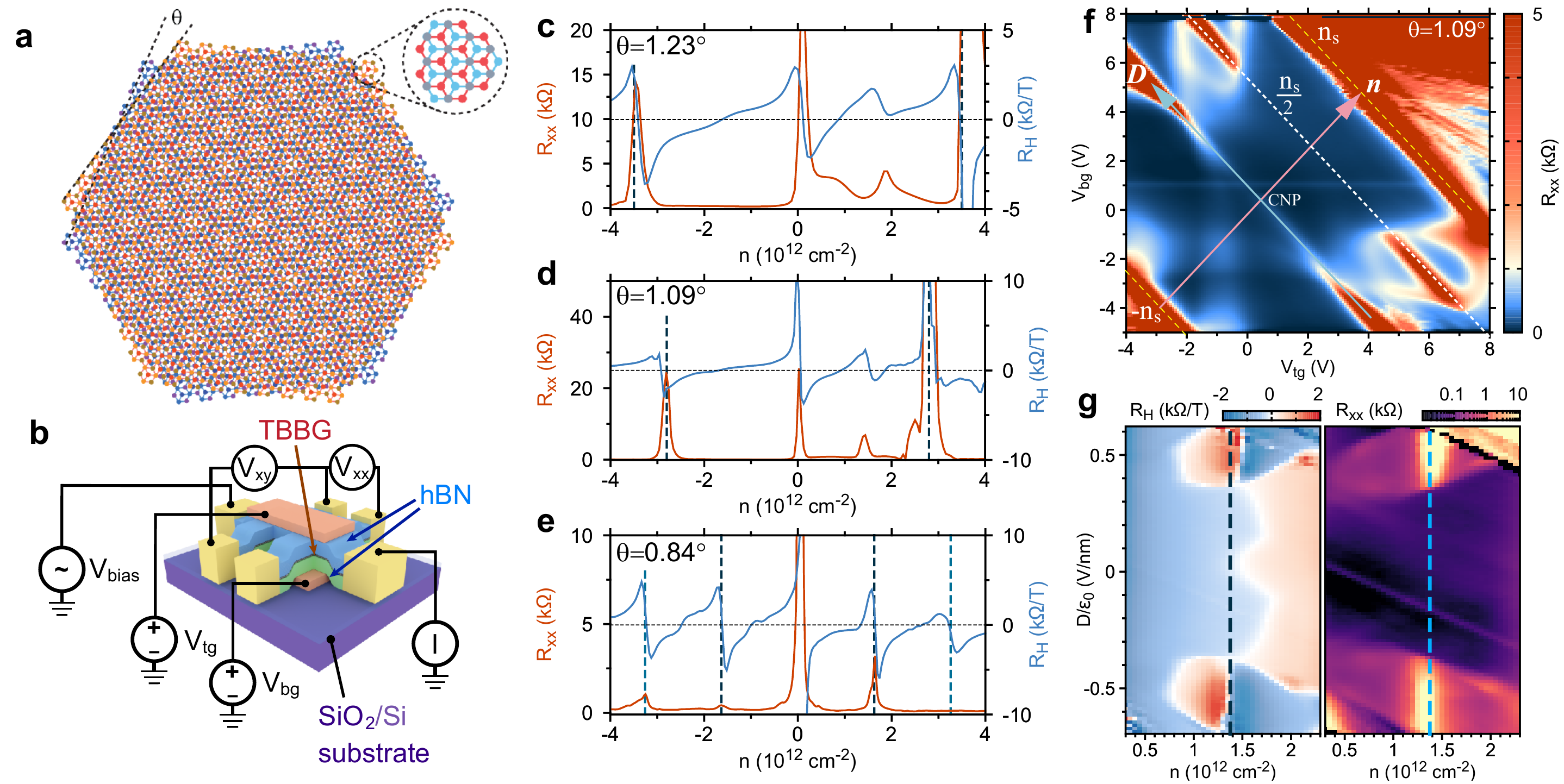}
\caption{Structure and transport characterization of twisted bilayer-bilayer graphene (TBBG). (a) TBBG consists of two sheets of Bernal-stacked bilayer graphene twisted at an angle $\theta$. (b) Schematic of a typical TBBG device with top and bottom gates and a Hall-bar geometry for transport measurements. (c-e) Measured longitudinal resistance $R_{xx}=V_{xx}/I$ and low-field Hall coefficient $R_H=\frac{\mathrm{d}}{\mathrm{d}B}(\frac{V_{xy}}{I})$ as functions of carrier density $n$ in three devices with twist angles $\theta=$\SI{1.23}{\degree}, \SI{1.09}{\degree} and \SI{0.84}{\degree} respectively. The vertical dashed lines denote multiples of the superlattice density $n_s$, where the peaking of $R_{xx}$ and sign changing of $R_H$ indicate the Fermi energy crosses a band edge of the superlattice bands. (f) Resistance of the \SI{1.09}{\degree} TBBG device versus both top gate and bottom gate voltages $V_{tg}$ and $V_{bg}$. The charge density $n$ and displacement field $D$ are related to the gate voltages by a linear transformation (see Methods). The superlattice densities $\pm n_s$ and the half-filling line at $n_s/2$ are indicated by dashed lines parallel to the $D$ axis. Correlated insulator states are observed at $n_s/2$ filling at finite displacement fields. (g) Map of low-field Hall coefficient $R_H$ (left panel) and resistance $R_{xx}$ (right panel) near the $n_s/2$ correlated states for the \SI{1.09}{\degree} TBBG device (the vertical dashed lines indicate density $n_s/2$) . We find that accompanying the onset of the correlated insulator states at $D/\varepsilon_0\approx\SI{+-0.35}{\volt\per\nano\meter}$, a new sign change of the Hall coefficient also emerges.}
\end{figure}

We fabricated high-mobility dual-gated TBBG devices with the reported `tear and stack' method\cite{cao_superlattice-induced_2016,kim_van_2016}, using exfoliated Bernal-stacked bilayer graphene instead of monolayer graphene. We measured the transport properties of six small-angle devices and here we focus on three of the devices with twist angles $\theta=$\SI{1.23}{\degree}, \SI{1.09}{\degree} and \SI{0.84}{\degree} respectively (see Extended Data Figure 1 for other devices). Figures 1c-e show the longitudinal resistance $R_{xx}$ and low-field Hall coefficient $R_H=\frac{dR_{xy}}{dB}$ versus charge density for these three devices at $T=\SI{4}{\kelvin}$. In a superlattice, the electronic band structure is folded in the mini Brillouin zone (MBZ), defined by the moir\'{e} periodicity\cite{bistritzer_moire_2011}. Each band in the MBZ can accommodate a total charge density $n_s=4/A$, where $A$ is the size of the moir\'{e} unit cell and the prefactor accounts for spin and valley degeneracies\cite{bistritzer_moire_2011,cao_superlattice-induced_2016,zhang_nearly_2019}. The experimental results show a sign change of the Hall coefficient $R_H$ at each multiple of $n_s$ (vertical dashed lines in Fig. 1c-e), indicating the switching of hole-like pockets to electron-like pockets, and peaks in $R_{xx}$, indicating the crossing of new band edges (for $\theta=\SI{0.84}{\degree}$, the band edges at $-n_s$ and $\pm2n_s$ may only have small gaps or may even be semi-metallic, and hence do not exhibit prominent peaks in $R_{xx}$. See Fig. 2d-f for band structure calculations). The sharpness of the peaks confirms that the devices show relatively low disorder and have well-defined twist angles. In the $\theta=\SI{1.23}{\degree}$ and $\theta=\SI{1.09}{\degree}$ devices, we observe signatures of newly formed gaps also at $n_s/2$ when an electrical displacement field is applied perpendicular to the device. The dual-gate device geometry allows to independently vary the total charge density $n=(c_{bg}V_{bg}+c_{tg}V_{tg})/e$, where $e$ is elementary charge and $c_{tg}$ and $c_{bg}$ are top and bottom gate capacitances per area respectively, and the electrical displacement field penetrating the sample $D=(c_{bg}V_{bg}-c_{tg}V_{tg})$, by applying differential gate voltages (see Methods for actual values). Fig. 1f shows the resistance map in the $V_{tg}$-$V_{bg}$ space for the $\theta=\SI{1.09}{\degree}$ device. The transformation between gate voltages and $(n, D)$ is given in the Methods section. At $D=0$, no insulating behavior other than the full-filling gaps at $\pm n_s$ is observed. However, when a displacement field $D$ is applied in either direction, an insulating state appears at $n_s/2$ for a range of $D$. This new insulator state induced by the displacement field can be further examined by measuring the Hall coefficient $R_H$ versus $n$ and $D$, as shown in Fig. 1g ($\theta=\SI{1.09}{\degree}$ device) together with $R_{xx}$ for comparison. At the onset of the insulating states at $D/\varepsilon_0\approx\SI{+-0.35}{\volt\per\nano\meter}$, where $\varepsilon_0$ is the vacuum permittivity, $R_H$ develops additional sign changes adjacent to the insulating states, suggesting the creation of new gaps by the displacement field. The insulating states disappear when $D/\varepsilon_0$ exceeds \SI{+-0.7}{\volt\per\nano\meter}. In both the $\theta=\SI{1.09}{\degree}$ device and the $\theta=\SI{1.23}{\degree}$ devices, we find signatures of the onset of a correlated state at $n=-n_s/2, D=0$, but no well-developed correlated state is observed (see Extended Data Figure 1). 

\begin{figure}
\includegraphics[width=\textwidth]{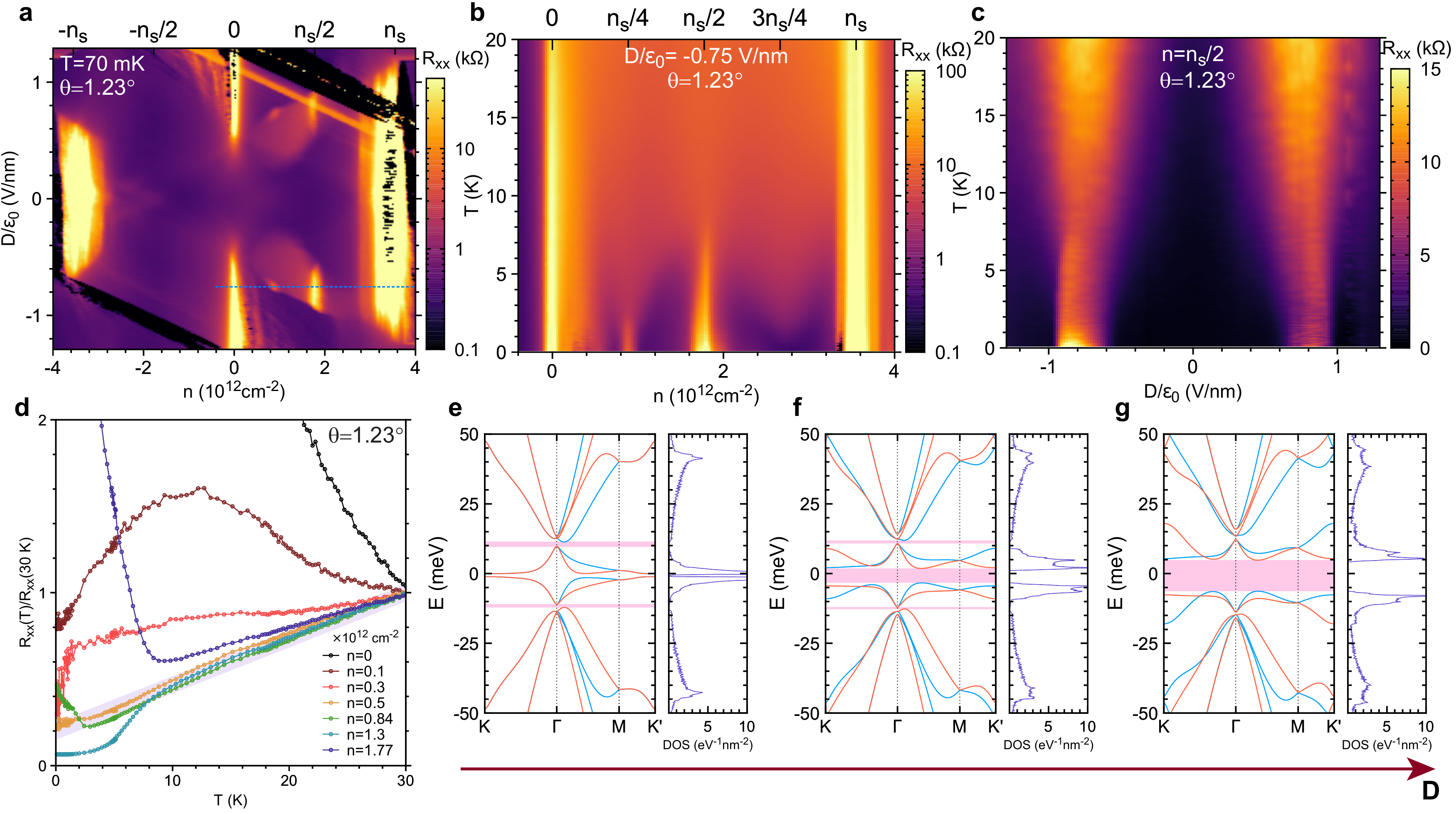}
\caption{Displacement field tunable correlated insulator states in TBBG. (a) Color plot of resistance versus charge density $n$ and displacement field $D$ ($\theta=\SI{1.23}{\degree}$ device, section 1, see methods). The blue dashed line cutting through the $D<0$ correlated state is the linecut along which (b) is taken (for $\theta=\SI{1.23}{\degree}$ device, section 2, see methods) (b) Resistance versus $n$ and $T$ at a fixed $D/\varepsilon_0=\SI{-0.75}{\volt\per\nano\meter}$. The correlated insulator states at $n_s/4$ and  $n_s/2$ are suppressed by raising the temperature. (c) Resistance at density $n_s/2$ versus displacement field and temperature. The resistance shows a maximum at approximately $D/\varepsilon_0=\SI{+-0.8}{V/nm}$, the region where the correlated insulator state is present. (d) Normalized resistance curves versus temperature at different densities between $0$ and $n_s/2\approx\SI{1.77e12}{\per\centi\meter\squared}$. Away from the charge neutrality point, all resistance curves show approximately linear R-T behaviour above \SI{10}{\kelvin}, with similar slopes (see Extended Data Figure 2). At around $n=\SI{1.3e12}{\per\centi\meter\squared}$, in proximity to the correlated state, an abrupt drop in resistance around \SI{6}{\kelvin} indicates the possible onset of superconductivity. (e-g) Calculated band structure (left panels) and density of states (right panels) for the $\theta=\SI{1.23}{\degree}$ TBBG device at (e) $\Delta V=0$, (f) $\Delta V=\SI{6}{\milli\volt}$ and (g) $\Delta V=\SI{12}{\milli\volt}$, where $\Delta V$ is the potential difference between adjacent graphene layers induced by the external displacement field (assumed to be the same between all layers). Single-particle bandgaps in the dispersion are highlighted by pink bars. }
\end{figure}

In the $\theta=\SI{1.23}{\degree}$ device, we observe a similar but more intricate hierarchy of tunable insulating states that stem from the interplay of correlations, the superlattice bands, and the magnetic field. Fig. 2a shows the $n$-$D$ resistance map for the $\theta=\SI{1.23}{\degree}$ TBBG device measured at $T=\SI{0.07}{\kelvin}$. Noticeably, as $|D|$ is increased the insulating state at charge neutrality $n=0$ strengthens in the same way as in Bernal-stacked bilayer graphene\cite{mccann_electronic_2013, oostinga_gate-induced_2008, zhang_direct_2009}, while the superlattice gaps at $\pm n_s$ weaken and eventually disappear (for $|D|/\varepsilon_0>\SI{1.2}{\volt\per\nano\meter}$ for the $+n_s$ insulating state and for $|D|/\varepsilon_0>\SI{0.7}{\volt\per\nano\meter}$ for the $-n_s$ insulating state). The band structures of TBBG in zero and finite external displacement fields calculated using a continuum approximation are shown in Fig. 2e-g (see Methods for details). It should be noted that, although TBBG has twice the number of graphene layers than TBG, the band counting is the same as TBG, \emph{i.e.} each band (spin/valley degenerate) accommodates 4 electrons per moir\'{e} unit cell. At zero displacement field, the gap at charge neutrality ($E=0$) is negligible while the superlattice gaps above and below the flat bands are visible. When the displacement field is increased, the charge neutrality gap quickly widens while the superlattice gaps become smaller and eventually vanish, in agreement with our data. 

At intermediate displacement fields around $D/\varepsilon_0=\SI{-0.75}{\volt\per\nano\meter}$, we observe not only insulator states at $n_s/2$ over a wide range of $D$, but also at $n_s/4$ over a smaller range (Fig. 2a). We attribute these states to a Mott-like mechanism similar to those observed in TBG, which is a result of the Coulomb repulsion of the electrons in the flat bands when each unit cell hosts exactly 1 or 2 electrons, corresponding to $n_s/4$ and $n_s/2$ fillings respectively. The $n_s/4$ state requires a finer tuning of $D$ to reveal, possibly due to the smaller gap size. This is evident from Fig. 2b, where we show the resistance versus $n$ and temperature $T$ with displacement field $D/\varepsilon_0$ fixed at \SI{-0.75}{V/nm}. Both states vanish as temperature rises. While the $n_s/2$ state persists up to approximately \SI{8}{\kelvin}, the $n_s/4$ disappears at less than \SI{3}{\kelvin}, indicating a smaller energy scale associated with the gap. Fig. 3c shows the resistance of the $n_s/2$ state versus displacement field and temperature. The 'optimal' displacement field is approximately \SI{+-0.8}{V/nm}. As temperature increases, the peak in $R_{xx}$ not only decreases in value but also broadens in $D$. As shown in Fig. 2d, at temperatures higher than \SI{10}{\kelvin} and away from the charge neutrality point, the transport is dominated by linear $R$-$T$ behaviour similar to that observed in TBG \cite{cao_unconventional_2018, yankowitz_tuning_2019, polshyn_phonon_2019}.  At a carrier density $n=\SI{1.3e12}{\per\centi\meter\squared}$ (lowest blue curve in Fig. 2d), the $R$-$T$ curve shows an abrupt drop at around \SI{6}{\kelvin}, saturating at a small, but finite, resistance. While the lowest resistance does not reach zero, possibly due to a series non-superconducting region due to disorder, the shoulder-like sudden drop in resistance near the correlated insulator state is reminiscent of similar superconducting transitions observed in TBG and rhombohedral trilayer graphene \cite{cao_unconventional_2018,yankowitz_tuning_2019,chen_signatures_2019,lu_superconductors_2019}, and our data suggest the proximity of our system to a superconducting phase around these charge density and displacement field values.

\begin{figure}
\includegraphics[width=\textwidth]{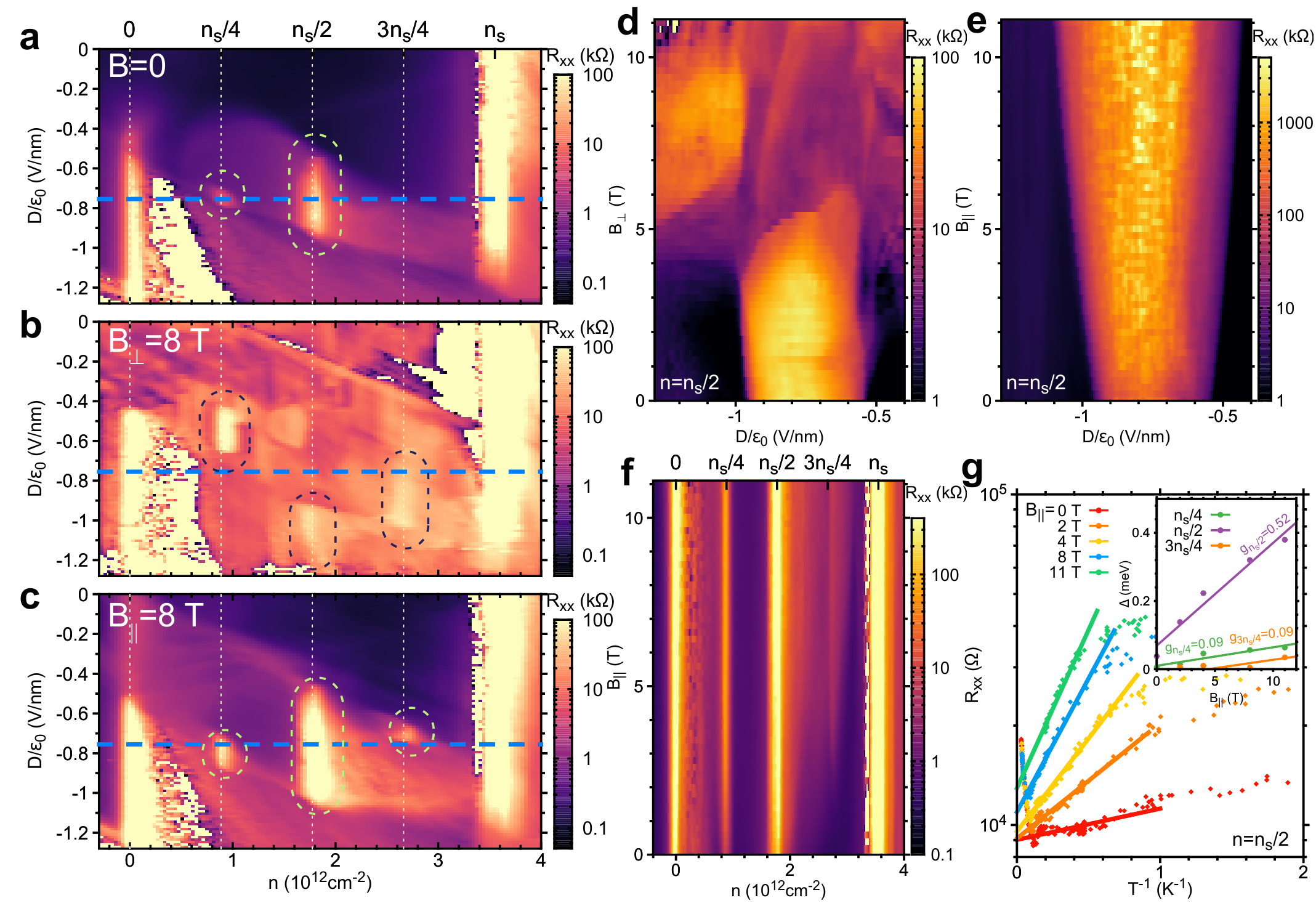}
\caption{Magnetic field response of the displacement-field-tunable correlated insulator states in TBBG. (a-c) Resistance plot for the $\theta=\SI{1.23}{\degree}$ TBBG device in magnetic fields of $B=0$, $B_\perp=\SI{8}{\tesla}$ perpendicular to the sample, and $B_\parallel=\SI{8}{\tesla}$ parallel to the sample, respectively. All measurements are taken at sample temperature  $T=\SI{0.07}{\kelvin}$. Various correlated states at integer electron fillings of the moir\'{e} unit cell are indicated by dashed circles. At zero field, only the $n_s/4$ and $n_s/2$ states appear around $|D|/\varepsilon_0=\SI{0.75}{\volt\per\nano\meter}$. In a perpendicular field of $B_\perp=\SI{8}{\tesla}$, the $n_s/4$ state shifts towards lower $|D|$, the $n_s/2$ state shifts towards higher $|D|$, and the $3n_s/4$ state also emerges. In a parallel field of \SI{8}{\tesla}, on the other hand, the position of the states barely shifts but their resistance increases monotonically. (d-e) Resistance at $n=n_s/2$ versus displacement field and magnetic field applied (d) perpendicular and (e) in-plane with respect to the device. While the correlated insulator state monotonically strengthens in $B_\parallel$, the perpendicular field induces a phase transition at around $B_\perp=$\SI{5}{\tesla}, where the correlated state abruptly shifts to higher $|D|$. (f) In-plane magnetic field response of the correlated insulators at $n_s/4$, $n_s/2$ and $3n_s/4$. (g) Resistance of the $n_s/2$ correlated state at an optimal $|D|$ (i.e. corresponding to maximum resistance), versus inverse temperature $T^{-1}$ for different in-plane magnetic fields. The solid lines are fit to an Arrenhius behaviour $\sim e^{-\Delta/2kT}$. Inset shows $\Delta$ obtained from the fitting for all three integer electron fillings versus $B_{\parallel}$. The $g$-factors extracted from these data are $0.52$ for the $n_s/2$ state, and $0.09$ for the $n_s/4$ and $3n_s/4$ states.}
\end{figure}

Figure 3 shows the response of the various correlated states to magnetic fields applied in either perpendicular or in-plane direction with respect to the sample plane. Figures 3a-c show the $n$-$D$ maps of the resistance for the $\theta=\SI{1.23}{\degree}$ device at $B=0$, $B_\perp=\SI{8}{\tesla}$ and $B_\parallel=\SI{8}{\tesla}$, respectively. The plots focus on densities from charge neutrality ($n=0$) to the first superlattice band edge ($n=n_s$). Fig 3a shows the band insulator states at $n=0$ and $n=n_s$, as well as the correlated states at $n_s/2$ and $n_s/4$ (encircled by dashed lines). At this zero magnetic field we do not observe an insulating state at 3 electrons per unit cell ($3n_s/4$ filling). Interestingly, at $B_\perp=\SI{8}{\tesla}$ (Fig. 3b), the correlated insulator states at $n_s/4$ and $n_s/2$ vanish at their original positions around $D/\varepsilon_0=\SI{-0.75}{V/nm}$ and reappear above and below their original positions at $B=0$, respectively, and a new correlated state appears now at $3n_s/4$. Fig. 3c shows that at $B_\parallel=\SI{8}{\tesla}$, correlated states are present at all integer fillings ($n_s/4$, $n_s/2$, and $3n_s/4$), and at approximately the same $D$ values as at $B=0$. Figure 3d shows the evolution of the $D$-position of the $n_s/2$ state as a function of $B_\perp$. It can be seen that at around $B_\perp=\SI{5}{\tesla}$, the correlated state shifts abruptly from $D/\varepsilon_0>\SI{-0.95}{V/nm}$ to $D/\varepsilon_0<\SI{-0.95}{V/nm}$. This peculiar behaviour might be attributed to a magnetic field induced spin-polarized insulator to valley-polarized metal phase transition at low $D$ (due to the competition between regular spin Zeeman and valley Zeeman effects), while at high $D$ we may have a corresponding metal to valley-polarized insulator transition. We note however that another possibility includes mixed spin-valley polarized states at low $B_\perp$ which turn into valley polarized states at high $B_\perp$.

On the other hand, the resistance for all the correlated insulator states increases smoothly and monotonically in a parallel magnetic field, as shown in Figs. 3e-f. In Figure 3g we analyze the evolution of the size of the correlated gaps with in-plane magnetic field by fitting the maximum resistance of the correlated states at "optimal" displacement fields with an Arrenhius behaviour $e^{-\Delta/2kT}$, where $k$ is the Boltzmann constant and $\Delta$ is the energy gap size. For all three correlated states at $n_s/4$, $n_s/2$ and $3n_s/4$ respectively, the gaps increase monotonically  by applying an in-plane magnetic field, with effective $g$-factors of 0.09, 0.52 and 0.09 respectively. By comparison, similar experiments on TBBG \cite{kim_ferromagnetic_2019}, have reported a $g$-factor close to 2 and attributed it to a possible purely spin-polarized ground state for the $n_s/2$ correlated state. While this argument does not apply to our data because the measured effective $g$-factors are much smaller than two, it is possible that the presence of disorder leads to a substantial underestimation of the gaps and their parallel magnetic field dependence. Alternatively, it is also possible that a mixed spin and valley polarization of the ground state has a reduced effective $g$-factor compared to a pure spin-polarized state. The monotonic increase in resistance with parallel magnetic field strongly suggest that the correlated insulator states have broken spin-rotation symmetry for all integer fillings, in stark contrast to the case of TBG, where the $n_s/2$ insulator state resistance decreases both with $B_\perp$ and $B_\parallel$ \cite{cao_correlated_2018, cao_unconventional_2018, yankowitz_tuning_2019}. Additional magnetic field response data is included in Extended Data Figure 3.

We note that all correlated insulator states in the $\theta=\SI{1.23}{\degree}$ TBBG device, whether at zero magnetic field or high magnetic fields, lie within the range $D/\varepsilon_0=$\SIrange{-1.2}{-0.4}{\volt\per\nano\meter}. Coincidentally, this is also the range where both the gap at charge neutrality ($n=0$) and the gap at the superlattice density ($n=n_s$) are well-developed, as can be seen in Figs. 3a-c. Based on this observation, we suggest that the displacement field tunability of the correlated states is tied to the modulation of the single-particle band gaps by the displacement field. When either gap at $n=0$ or $n=n_s$ is absent, the thermally excited or disorder scattered carriers from the upper or lower band would suppress the ordering of the electrons and hence the correlated states. Further theoretical work is needed to reveal the detailed structure of the displacement field dependence of the correlated states. 

\begin{figure}
\includegraphics[width=\textwidth]{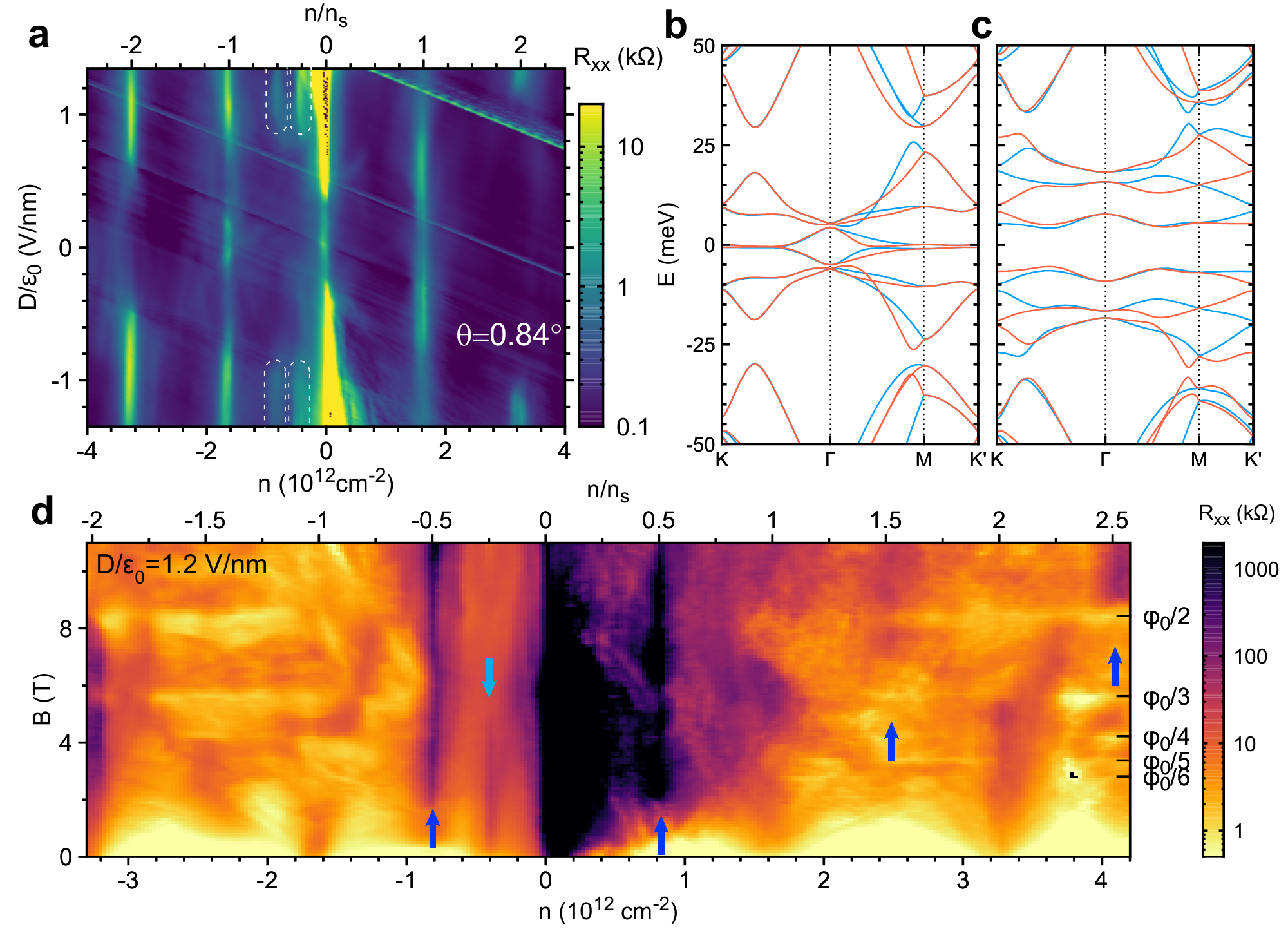}
\caption{Correlated insulator states in a multi flat band system. (a) Resistance map of a $\theta=\SI{0.84}{\degree}$ TBBG device. The top axis is the charge density $n$ normalized to the superlattice density $n_s$. Besides the $D$-tunable gaps at multiples of $n_s$, we find signatures of correlated states at $n/n_s=-\frac{1}{2}, -\frac{1}{4}$ for $|D|/\varepsilon_0 > \SI{0.8}{\volt\per\nano\meter}$, which are indicated by dashed circles. (b-c) Calculated band structure of $\theta=\SI{0.84}{\degree}$ TBBG (b) without interlayer potential and (c) with interlayer potential $\Delta V=\SI{18}{\milli\volt}$. Near charge neutrality, within a \SI{50}{\milli\electronvolt} window, there are in total six sets of flat bands spanning densities $-3n_s$ to $3n_s$. Upon applying a displacement field, these bands are further flattened and separated from each other, which makes them more prone to giving rise to correlated physics at each half-filling. (d) Resistance as a function of charge density and perpendicular magnetic field $B$. We observe strong correlated states at $n/n_s=-\frac{1}{2}$ and $\frac{1}{2}$, and weaker states at $\frac{3}{2}$ and $\frac{5}{2}$ fillings. All the half-filling correlated states (indicated by dark blue arrows) are enhanced by the application of a magnetic field. The quarter-filling state at $n/n_s=-\frac{1}{4}$ (indicated by the light blue arrow) is however suppressed by the magnetic field. Due to the formation of a superlattice, we also observe Hofstadter butterfly related conductive features (bright horizontal features, see right axis) when $B$ is such that the magnetic flux in each unit cell is equal to $\varphi_0/2, \varphi_0/3, \varphi_0/4, \ldots$, where $\varphi_0=h/e$ is the flux quantum flux. }
\end{figure}

We have also investigated the effect of substantially reducing the twist angle. Unlike the case of TBG, further reduction of the twist angle of TBBG to \SI{0.84}{\degree} results not only in one, but actually three pairs of flat bands, separated from other bands by band gaps (Fig. 4b). Application of an electrical displacement field further flattens these bands and separate them from each other (Fig. 4c). This would imply that all electrons within a density range $-3n_s$ to $+3n_s$ might experience strong Coulomb interactions, and their correlations can get further enhanced by applying a displacement field. These predictions from band theory are indeed consistent with our experimental observations. In Fig. 4a where we show the resistance map of the $\theta=\SI{0.84}{\degree}$ TBBG device versus $n$ and $D$, we find weak signatures of $-n_s/2$ and $-n_s/4$ correlated insulator states only at high displacement fields $|D|/\varepsilon_0 > \SI{0.8}{\volt\per\nano\meter}$ (encircled by white dashed lines). Noticeably, the full-filling gaps at $\pm n_s$ and $\pm 2n_s$ are also tunable by the displacement field to different extents. Figure 4d shows the effect of applying a perpendicular magnetic field, where more correlated states are revealed similar to the spin-polarized states in the $\theta=\SI{1.23}{\degree}$ device. The strongest correlated states are observed at $n/n_s=\pm \frac{1}{2}$, and weaker states are visible at $n/n_s=\frac{3}{2}$ and $\frac{5}{2}$ in high magnetic fields. This is consistent with the fact that in Fig. 4c, the pair of bands closer to charge neutrality is flatter than the other two pairs of bands farther away from the charge neutrality. All correlated states at half-filling of these flat bands are enhanced by $B$, indicating at least partial spin-polarized ordering. The resistance of the quarter-filling state at $n/n_s=-\frac{1}{4}$, on the other hand, does not increase with perpendicular field, possibly due to a more intricate interplay between orbital and spin effects.

Our results show that TBBG exhibits a rich spectrum of correlated phases tunable by twist angle, electric displacement field and magnetic field, paving the way to further studies of strongly correlated physics and topology in multi-flat band systems\cite{zhang_nearly_2019}.

\bibliographystyle{MyStyle}
\bibliography{references} 

\section*{Acknowledgements}
We acknowledge helpful discussions with S. Todadri, L. Fu, P. Kim, X. Liu, S. Fang and E. Kaxiras. 

\section*{Author Contributions}
Y.C., D.R-L., O.R-B. and J.M.P fabricated samples and performed transport measurements. Y.C., D.R-L., O.R-B., J.M.P and P.J-H. performed data analysis. K.W. and T.T. provided h-BN samples. Y.C., D.R-L., O.R-B., J.M.P. and P.J-H. wrote the manuscript with input from all co-authors.

\section*{Author Information}
The authors declare no competing financial interest.

\newpage

\section*{\Large{Supplementary Information}}

\renewcommand{\figurename}{Extended Data Figure}
\setcounter{figure}{0}

\section{Fabrication}

The reported devices are fabricated with two sheets of Bernal-stacked bilayer graphene (BLG) and encapsulated by two hBN flakes. Both BLG and hBN were exfoliated on SiO\textsubscript{2}/Si substrates, and the thickness and quality of the flakes were confirmed with optical microscopy and atomic force microscopy. A modified polymer-based dry pick-up technique was used for the fabrication of the heterostructures. A poly(bisphenol A carbonate)(PC)/polydimethylsiloxane(PMMA) layer on a glass slide was positioned in the micro-positioning stage to first pick up an hBN flake at \SI{100}{\celsius}. The van der Waals interaction between the hBN and BLG then allowed us to tear the BLG flake, which was then rotated at a desired angle and stacked at room temperature. The resulting hBN/BLG/BLG heterostructure was released on another hBN flake on a Pd/Au back gate that was pre-heated to \SI{170}{\celsius}, using a hot transfer technique \cite{pizzocchero_hot_2016, purdie_cleaning_2018}. The desired geometry of the four-probe devices was achieved with electron beam lithography and reactive ion etching. The electrical contacts and top gate were deposited by thermal evaporation of Cr/Au, making edge contacts to the encapsulated graphene\cite{wang_one-dimensional_2013}. 

\section{Measurements}

Electronic transport measurements were performed in a dilution refrigerator with a superconducting magnet, with a base electronic temperature of \SI{70}{\milli\kelvin}. The data were obtained with low-frequency lock-in techniques. We measured the current through the sample amplified by \SI{1e7}{\volt\per\ampere} and the four-probe voltage amplified by $1000$, using SR-830 lock-in amplifiers that were all synchronized to a frequency of \SIrange{1}{10}{\hertz}.

\section{List of Measured TBBG Devices}

Following the definition given in the main text and accounting for offsets in the gate voltages due to impurity doping, $n$ and $D$ are related to the top and bottom gate voltages $V_{tg}$ and $V_{bg}$ by
\begin{align}
n &= [c_{tg}(V_{tg}-V_{tg,0}) + c_{bg}(V_{bg}-V_{bg,0})]/e, \\
D &= [-c_{tg}(V_{tg}-V_{tg,0}) + c_{bg}(V_{bg}-V_{bg,0})].
\end{align}

Table \ref{tab:s1} lists the twist angles and parameters $c_{tg}$, $c_{bg}$, $V_{tg, 0}$, $V_{bg, 0}$, and $n_s$ for all devices discussed in this work, including those shown in the Extended Data Figures. These parameters are estimated to satisfy that all diagonal features in the $V_{tg}$-$V_{bg}$ maps are rotated to be vertical in the corresponding $n$-$D$ maps, and the features should be symmetrical with respect to $D$ after the transformation. 

\begin{table}[!ht]
    \centering
    \begin{tabular}{|c|c|c|c|c|c|}
         \hline
         $\theta$(\si{\degree}) & $c_{tg}$ (\si{\farad\per\meter\squared}) & $c_{bg}$ (\si{\farad\per\meter\squared}) & $V_{tg, 0}$ (\si{V}) & $V_{bg, 0}$ (\si{V}) & $n_s$ (\si{cm^{-2}}) \\\hline
         1.09 & \num{6.63e-4} & \num{5.02e-4} & 0.30 & 0.58 & \num{2.75e12}\\\hline
         1.23 & \num{1.06e-3} & \num{7.14e-4} & 0.41 & -0.04 & \num{3.55e12}\\\hline
         1.23 & \num{1.06e-3} & \num{7.14e-4} & 0.41 & -0.04 & \num{3.55e12}\\\hline
         0.84 & \num{6.87e-4} & \num{6.38e-4} & 0.06 & 0.08 & \num{1.65e12}\\\hline
         0.79 & \num{1.06e-3} & \num{3.57e-4} & 0.18 & 0.67 & \num{1.45e12}\\\hline
         1.09(*) & \num{1.03e-3} & \num{5.12e-4} & 0.28 & 0.45 &
         \num{2.75e12}\\\hline
         
    \end{tabular}
    \caption{List of TBBG devices discussed in the main text and Extended Data Figures. The last device is marked with an asterisk to differentiate it from the first device, which happens to have the same twist angle but it is a totally independent device fabricated on a separate chip, and will be hereinafter referred to with the asterisk to avoid confusion.}
    \label{tab:s1}
\end{table}

In Extended Data Figures 1a-f, we show $V_{tg}$-$V_{bg}$ resistance maps for all six TBBG devices we measured. Extended Data Figures 1c-d are measured in the same TBBG sample, but in different sample regions that are approximately \SI{27}{\micro\meter} apart (sections 1 and 2, respectively). Both regions have identical parameters (hence the two identical rows in table I), with the same twist angle $\theta=\SI{1.23}{\degree}$, and also nearly identical transport characteristics. The two sections are electrically disconnected via etching, but the extracted twist angles from the data have a difference of less than \SI{0.01}{\degree}, suggesting very uniform twist angles across this entire sample.

In almost all TBBG samples, we noticed a peculiar cross-like pattern around $(n,D) = (-n_s/2, 0)$, \emph{i.e.} near p-side half-filling of the superlattice band. This is especially apparent in the \SI{1.09}{\degree} and \SI{1.23}{\degree} devices, which are highlighted in Extended Data Figures 1g-h. The p-side band does not exhibit a strong $D$-tunable correlated state as elaborated in the main text, possibly due to the larger bandwidth compared to its n-side counterpart. This cross-like pattern might represent an onset of correlated behaviour near half-filling of the band. Further experimental work and theoretical insights are needed to understand this phenomenon.

\section{Linear $R$-$T$ behaviour}

Extended Data Figure 2 shows the resistance versus temperature behaviour, at different densities, observed across several small-angle TBBG devices. In the \SI{1.23}{\degree} device, we find approximately linear $R$-$T$ behaviour above \SI{10}{\kelvin} for densities ranging from \SIrange{0.5e12}{2.5e12}{\per\centi\meter\squared}, encompassing the $n_s/2$ correlated state. The resistance slope in this range of densities does not vary very substantially, ranging from \SIrange{210}{350}{\ohm\per\kelvin}. Since all our devices have length-to-width ratios close to one, these slope values are therefore close to those reported in TBG\cite{cao_strange_2019,polshyn_phonon_2019}. In stark contrast, the resistance behaviour in the hole-doping side ($n<0$), as shown in Extended Data Figure 2b, shows qualitatively different behavior: it does not show linear $R-T$ characteristics, at least up to \SI{30}{\kelvin}, and the resistance value is about an order of magnitude smaller than on the electron-doping side. These data are consistent with the picture that the electron-doping band is flatter than the hole-doping band, therefore exhibiting more pronounced correlated phenomena, examples being the $n_s/2$ insulator state, linear resistance-temperature behaviour, and possibly superconductivity. Extended Data Figure 2c shows $R$-$T$ curves close to the $n_s/2$ state. 

The data for the \SI{1.09}{\degree} device shows a similar trend of linear $R$-$T$ behaviour starting around \SI{5} to \SI{10}{\kelvin}, as shown in Extended Data Figure 2d. 

In the \SI{0.84}{\degree} device, we find a very different behaviour. Although there is a region of approximately linear $R$-$T$ behaviour at all densities, except at multiples of $n_s$, the resistance slope is now strongly dependent on the charge density $n$. The slope approximately follows a power law $dR_{xx}/dT \propto n^a$ where $a\approx -1.77$ (see inset).

\section{Theoretical Methods}

The band structures shown in the main text are calculated using a continuum model based on the original continuum model for TBG \cite{bistritzer_moire_2011, lopes_dos_santos_continuum_2012}, which qualitatively captures most of the important features of the bands in TBBG including displacement field dependence. To lowest order, the continuum model of twisted graphene superlattices is built on the approximation that the interlayer coupling between the A/B sublattice of one layer and
the A/B sublattice of the other layer has a sinusoidal variation over the periodicity of the moir\'{e} pattern. For the three possible directions of interlayer connections between the wave vectors in the Brillouin zone, there are three connection
matrices, 
\begin{align}
H_1 &= w\begin{pmatrix}1 & 1\\1 & 1\end{pmatrix}\\
H_2 &= w\begin{pmatrix}\omega^2 & 1\\\omega & \omega^2\end{pmatrix}\\
H_3 &= w\begin{pmatrix}\omega & 1\\\omega^2 & \omega\end{pmatrix}
\end{align}
where $w$ is the interlayer hopping energy and
$\omega=e^{2\pi i/3}$. $H_{i, \alpha\beta}$, with $\alpha,\beta=A,B$ represents the hopping between sublattice $\alpha$ in the first layer to sublattice $\beta$ in the second layer, with momentum transfer determined by $i$ (see \cite{bistritzer_moire_2011} for definition). Note that in this gauge choice, the origin of rotation is chosen where the B sublattice of the first layer coincides with the A sublattice of the second layer, so that the $H_{i,BA}$ component has zero phase while the other terms acquire phases. A different gauge choice is equivalent to an interlayer translation, which has been shown to have a negligible effect in the case of small twist angles \cite{bistritzer_moire_2011, lopes_dos_santos_continuum_2012}.

To extend this formulation to TBBG, we add a simplified bilayer graphene Hamiltonian,
\begin{align}
H_b = \begin{pmatrix}0 & 0\\w_b & 0\end{pmatrix} 
\end{align}
between the non-twisted layers. The momentum transfer is zero since the bilayer is not twisted and the coupling is constant over the moir\'{e} unit cell. For simplicity, we only consider the `dimer' coupling in the bilayer, neglecting second-nearest-neighbor hopping terms and trigonal warping terms. The two bilayers in TBBG (layer 1-2 and layer 3-4) have the same stacking order, \emph{i.e.} for zero twist angle the total stacking would be `ABAB' instead of `ABBA'. In the calculations used in the main text, we used parameters $w=\SI{0.1}{\electronvolt}$ and $w_b=\SI{0.4}{\electronvolt}$, so that when either parameter is turned off we obtain either the two non-interacting bilayer graphene ($w=0$) or the non-interacting TBG and two monolayer graphene ($w_b=0$). 

\section{Additional Magnetic Field Response Data}

Extended Data Figures 3a-b show the resistance in log scale versus inverse temperature at $n_s/4$ and $3n_s/4$ in the \SI{1.23}{\degree} device ($D$ chosen to maximize resistance), similar to Fig. 3g in the main text. The energy gaps $\Delta$ fitted by the Arrenhius formula $e^{-\Delta/2kT}$ for $n_s/4$ and $3n_s/4$ are shown in the inset of Fig. 3g. Extended Data Figure 3c-d show the response of correlated states at $n_s/4$ and $3n_s/4$ in a perpendicular magnetic field, similar to Fig. 3d. For the $n_s/4$ state, we again find a signature of a phase transition at $D/\varepsilon_0=\SI{-0.72}{V/nm}$, manifesting as a shift of $D$-location of the correlated insulator as $B_\perp$ exceeds \SI{5}{\tesla}. The $3n_s/4$ state shows an overall monotonic increase of resistance and exhibits no shift of the position in $D$. In an in-plane field on the other hand, as shown in Extended Data Figure 3e-f, both quarter filling states show a monotonic enhancement as $B_\parallel$ is increased.

\newpage

\begin{figure}
    \centering
    \includegraphics[width=\textwidth]{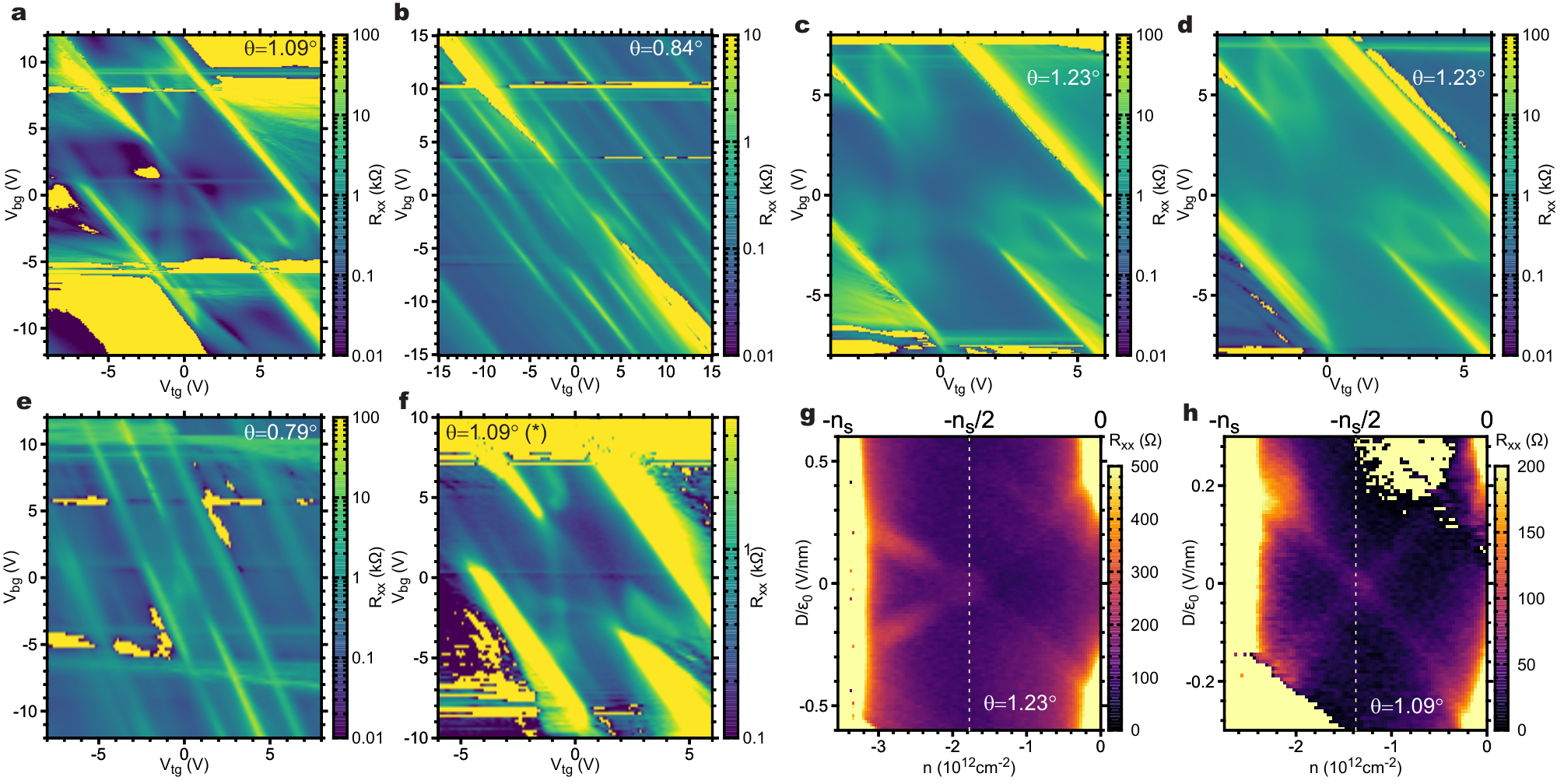}
    \caption{\textbf{$V_{tg}$-$V_{bg}$ resistance maps of various TBBG devices.} (a-f) Resistance versus $V_{tg}$ and $V_{bg}$ for five TBBG devices we measured so far. (g-h) Cross-like feature near $-n_s/2$ in TBBG samples, which might signal the onset of a correlated state.}
    \label{fig:edf1}
\end{figure}

\begin{figure}
    \centering
    \includegraphics[width=\textwidth]{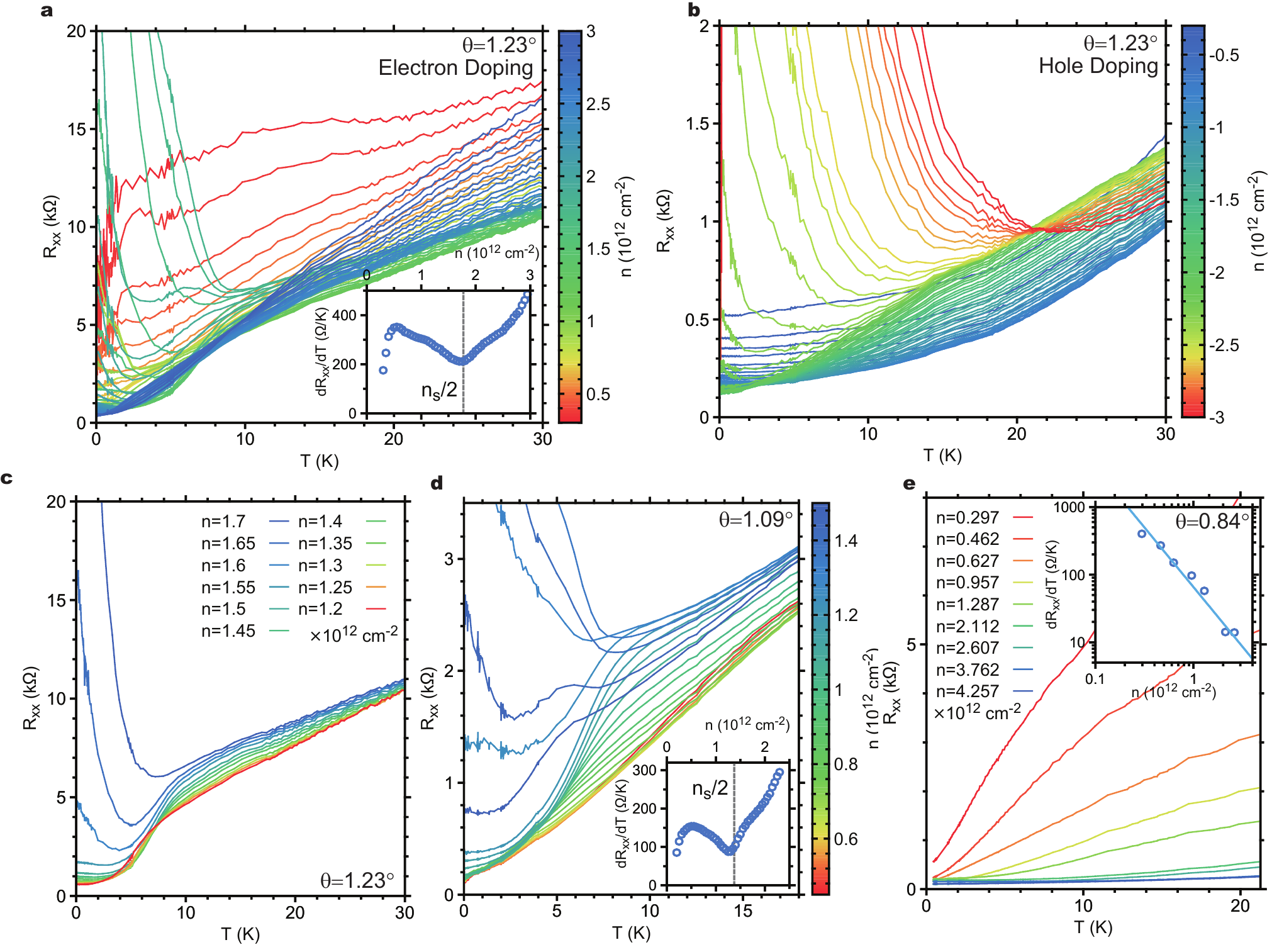}
    \caption{\textbf{Linear resistance-temperature behaviours in TBBG.} (a-b) Resistance versus temperature curves at different charge densities in the \SI{1.23}{\degree} sample. Inset of (a) shows the slope $dR_{xx}/dT$ of the linear $R$-$T$ behaviour as a function of $n$ for $T>\SI{10}{\kelvin}$. (c) Selected $R$-$T$ curves near $n_s/2$ from (a). (d) Similar linear $R$-$T$ behaviour in the \SI{1.09}{\degree} device. Inset shows the slope $dR_{xx}/dT$. (e) Density-dependent linear $R$-$T$ behaviour in the \SI{0.84}{\degree} deice. Inset shows the slope $dR_{xx}/dT$ versus $n$ in log-log scale. The slope is proportional to $n$ to the power of \num{-1.77}.}
    \label{fig:edf2}
\end{figure}

\begin{figure}
    \centering
    \includegraphics[width=\textwidth]{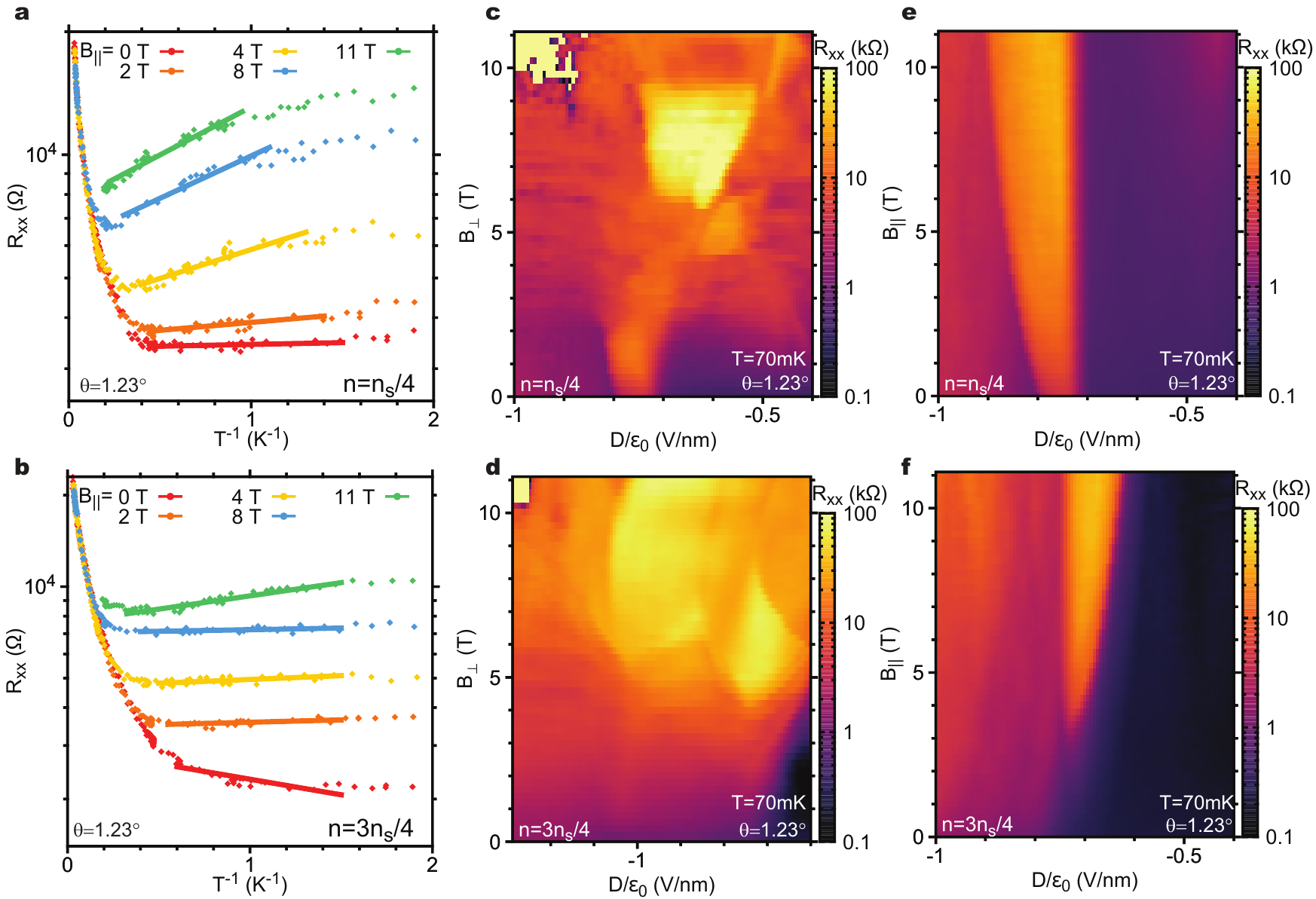}
    \caption{\textbf{Additional magnetic field response of TBBG devices.} (a-b) Resistance in log scale versus inverse temperature for $n_s/4$ and $3n_s/4$ states in the \SI{1.23}{\degree} sample. (c-f) Response of the $n_s/4$ and $3n_s/4$ states in either perpendicular magnetic field or in-plane magnetic field.}
    \label{fig:edf3}
\end{figure}

\end{document}